# Docs are ROCs: a simple off-the-shelf approach for estimating average human performance in diagnostic studies

Dr Luke Oakden-Rayner, Dr Lyle Palmer

The Australian Institute for Machine Learning, The University of Adelaide, Australia.
luke.oakden-rayner@adelaide.edu.au
lyle.palmer@adelaide.edu.au

## Introduction

Sensitivity and specificity are among the most common and important metrics in diagnostic medical research, favoured for their ability to summarise both false positive and false negative errors[1]. Their invariance to disease prevalence allows for a direct numerical comparison of performance across tests or sites with different rates of disease. Unsurprisingly, these metrics are widely reported in medical artificial intelligence (AI) studies, particularly where human performance is compared to that of AI models in multi-reader multi-case (MRMC) study designs. Given the variability in performance between readers on any given set of cases, multiple human readers are required to estimate the range of "average" human performance, but there is no well-motivated consensus on how to perform this averaging operation. A recent systematic review[2] noted that 'naive' averages of human sensitivity and/or specificity, or other metrics derived from these values, were reported in at least 70% of publications that compared human performance to AI models, a practice which is highly problematic.

The use of sensitivity and specificity to describe the discriminative performance of individual tests or readers is appropriate, but averaging these highly correlated metrics independently of each other is strongly discouraged in other, more methodologically mature domains such as the meta-analysis of diagnostic test accuracy studies. For example in "Guidelines for Meta-Analyses Evaluating Diagnostic Tests"[3] the authors write "In general, estimating mean sensitivity and specificity separately underestimates test accuracy". Gatsonis and Paliwal[4] even go as far as to say "the use of simple or weighted averages of sensitivity and specificity to draw statistical conclusions is not methodologically defensible." Similarly, the Cochrane handbook recommends these metrics be addressed together rather than in isolation when summarising the accuracy of a diagnostic test[5].

Despite these recommendations from reputable authors and bodies, the independent pooling of sensitivity and specificity (or use of similar pooled metrics such as the average F1 score and average accuracy) remain popular in the medical AI literature. Unfortunately, not only do these methods consistently underestimate human diagnostic performance, but they can bias the experimental conclusions because the performance of the AI models are not similarly underestimated.

An alternative to the various pooled metrics is to treat the estimation of human performance as a bivariate modelling problem, operating on the justified assumption that sensitivity and specificity are correlated across readers. This approach has become the mainstay of the meta-analysis of diagnostic test accuracy across the last 50 years, supported by an extensive body of literature on the development and validation of meta-analytic models. Indeed, meta-analysis is considered the highest level of experimental evidence in clinical medicine [6], in part because of the robustness of these techniques.

Meta-analysis for diagnostic test accuracy studies involves the production of summary receiver operating characteristic (SROC) curves. The specific methods to do this are well covered in other publications[7–10], but it is worth noting that these techniques are well understood and validated in the biostatistics community, and that software implementations of these methods are widely available in most common programming languages. Typically all that is needed is the 2x2 contingency table for each reader and the software can do the rest.

**Fixed effects vs random effects models**

Briefly, there are two main families of models used for meta-analysis and SROC curve development; the fixed-effects and random-effects models. In simple terms the fixed effect models assume that the only difference between tests (in this case, the readers) is due to a single source of variation; that it is the choice of readers alone that contributes to these differences. In random-effects models, sometimes called hierarchical or two-level models, an estimation of further test heterogeneity is included. In the setting of reader studies common further sources of test heterogeneity include intra-user variability and the assessment of different cases by each reader.

In general, random-effects models are recommended for the meta-analysis of diagnostic test accuracy studies[5]. Given the multiple sources of heterogeneity in MRMC studies, this recommendation appears appropriate in this context as well.

**Related literature**

In the case where the outcome is binary, a point in ROC space will be produced representing the sensitivity and specificity derived from a simple 2x2 table. However, in order to derive ROC curves at the level of individual readers, the MRMC literature has been strongly focused on the use of ordinal scoring systems such as those used in mammography[11]. In the scenario where a diagnostic score contains at least 5 levels it is reasonable to produce ROC curves for individual readers and then average these curves themselves, summarising the performance across readers[12]. These methods cannot be used when the clinical diagnosis is made with a binary response (e.g., "yes there is cancer" vs "no there is not"), and attempts to extend these methods to binary data[13] have not seen widespread uptake. In fact, it has become accepted practice to shoehorn ordinal scoring systems into tasks normally reported with binary responses (e.g., applying a 100 point scoring system to lung nodule detection[14], despite the fact that clinical radiologists only ever report that nodules are present or absent). This approach has even been tacitly endorsed by regulatory bodies[15], but is critically limited by its failure to reproduce clinical practice, raising significant concerns about the clinical relevance of this testing and the possibility of misleading laboratory effects[16].

In the medical AI literature, Rajpurkar et al used a constrained spline approach to summarise performance by estimating AUC[17]. This method assumes a symmetrical ROC curve (which is an uncommon distribution of readers in clinical practice), and produces confidence intervals with a bootstrap across cases alone, therefore underestimating the standard errors of the AUC by failing to incorporate the variability across readers.

**Why do SROC analysis of multi-reader studies?**
Aside from the already stated improvements in accuracy and methodological defensibility, SROC analysis has a number of attractive features compared to other commonly used methods.

First, it allows for the estimation of a single metric (the area under the SROC curve, also known as the SROC AUC) which summarises the discriminative ability of readers. Comparison between reader groups or readers and AI models is significantly simplified compared to separate consideration of sensitivity and specificity or similar metrics.

Second, it allows us to produce valid confidence intervals. When sensitivity and specificity are pooled separately, the confidence intervals are almost always calculated using the number of cases but *not* the number of readers. SROC analysis automatically takes both elements of variation into account. Importantly in common experimental scenarios (where $n_{Observers}$ <10) the number of readers contributes strongly to the estimation of variation.

Third, it avoids the need to select an arbitrary or unnatural (i.e., one that will never occur in clinical practice) operating point. If we consider that the position of a human reader along an SROC curve is related to their "aggressiveness" or risk-aversion, then these quantities are not fixed, either between readers or for individuals. SROC analysis allows for more control of the selection of an operating point if this is needed, and allows comparisons without operating point selection if this is more appropriate.

Fourth, SROC analysis allows for visual presentation of results in a way that is easy to interpret. Side-by-side ROC curves are understandable at a glance while conveying a great deal of information about the discriminative performance of each decision maker, and the ability to plot confidence intervals allows for a useful visual summary of an experiment.

Fifth and finally, SROC analysis can allow for easy comparison of subsets of readers. Many studies have included both expert and non-expert readers, and presentation of these results can be difficult. Single summary points (pooled sensitivity and specificity) are unjustified, but colour-coding of all the readers can be visually overwhelming if $n_{Observers}$ is high. Producing SROC curves for each subset can allow for easy comparisons between groups, and comparisons of SROC AUC values are well motivated (particularly given the differing number of readers and different variance between these subgroups).

## Methods

We present examples of this meta-analytic approach applied to a variety of heavily cited reports in the medical AI literature, re-evaluating the presented ROC curves and primary comparisons. For the majority of these papers, the data from these studies have been reproduced from the published figures (i.e., sensitivity and specificity were "eyeballed" for each reader), although Tschandl et al. provided the raw reader data for their experiments[18].

All statistical analysis was performed in R v3.6.2[19] . SROC analysis was undertaken with the mada package v0.5.8[20], using the proportional hazards model described by Holling et al[9].

## Results

**Dermatologist-level classification of skin cancer with deep neural networks**

Esteva et al[21] described a deep learning model trained to distinguish melanoma from non-melanomatous skin lesions, comparing the performance of the model against 22 dermatologists asked to decide if a skin lesion requires biopsy.

Esteva et al reported the average performance of the dermatologists by pooling sensitivity and specificity independently. This "average dermatologist" point was inside the ROC curve for the AI model. This figure was accompanied by the statement that the "CNN outperforms any dermatologist whose sensitivity and specificity point falls below the blue curve of the CNN", although no specific statement was made about the "average" dermatologist.

In figure 1 we apply a random-effects model meta-analysis of the performance of the dermatologists, showing the benefits of treating sensitivity and specificity as correlated values. The average point is inside the summary ROC curve, and in fact is at the limit of the 95% confidence interval. The SROC curve appears to better reflect the desired goal of describing an average dermatologist. With the curved distribution of the model as a guide, only 4 out of 22 dermatologists are "worse" than the average sens/spec point.

This approach not only produces a more justified summary of human performance, but the area under the SROC curve is directly comparable to the AUC of the AI model. The reported AUC of the AI model (0.94, CI not provided) is compared to the dermatologists (SROC AUC = 0.97, 95% CI 0.96 - 0.98).

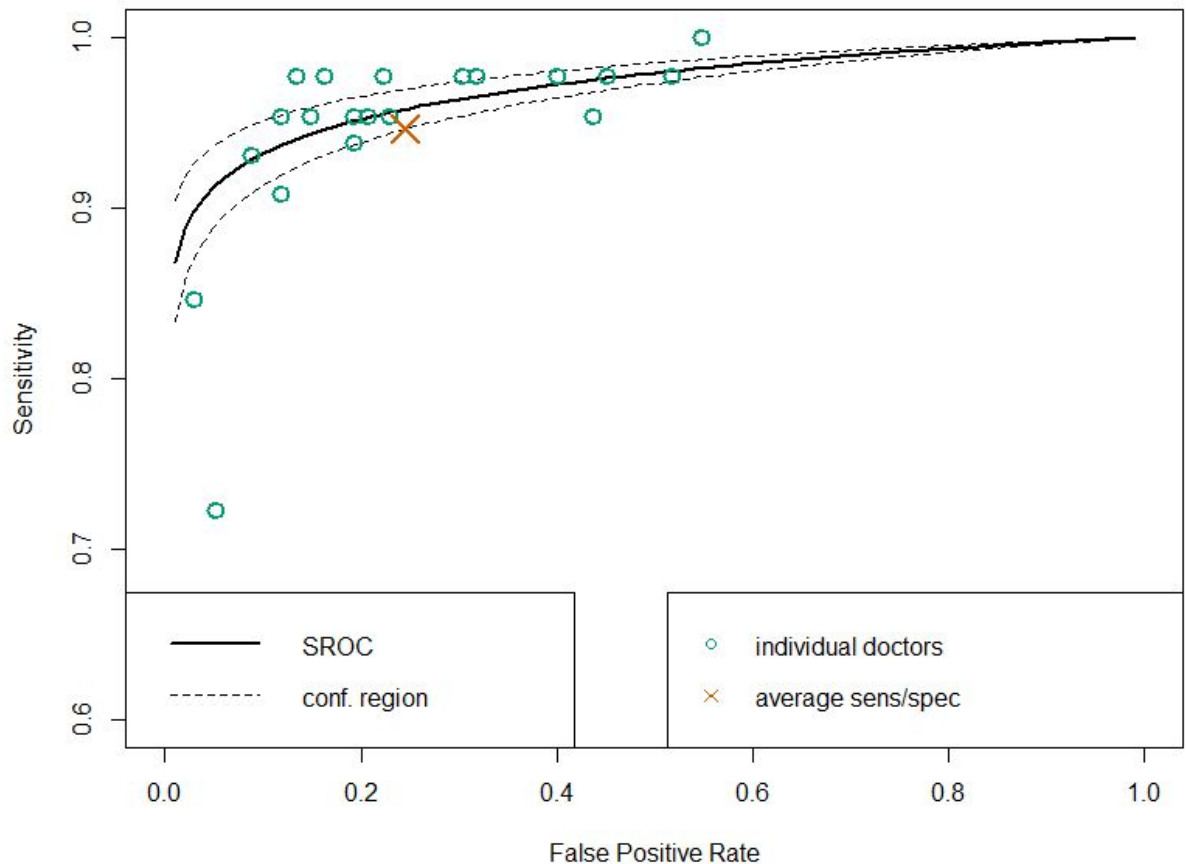

Figure 1: SROC analysis of Esteva et al[21] using a random effects model, demonstrating the individual performance of doctors (green circles), the average sensitivity and specificity of doctors (orange cross) and the SROC curve (black line) with associated 95% confidence region (dotted lines).

**Deep-learning-assisted diagnosis for knee magnetic resonance imaging: Development and retrospective validation of MRNet**

Bien and Rajpurkar et al[22] reported the comparison of a deep learning model against radiologists and orthopedic surgeons at the detection of meniscal tears, ACL tears, and combined for any abnormality. They reported the average performance of the dermatologists by pooling sensitivity and specificity independently.

In figure 2 we apply a random-effects meta-analysis to the performance of the clinical readers at the "any abnormality" task. Once again the "average" reader is below the SROC curve, and the SROC curve appears to be a more fair reflection of average reader performance.

The authors report that there was no significant difference between doctors and the AI model performance, albeit they allow for the fact that both the number of readers and number of cases are quite low leading to wide confidence intervals. In our approach, the AI model AUC of 0.937 (95% CI 0.895, 0.980) can be directly compared to the SROC AUC of 0.953 (95% CI 0.937, 0.969), which supports the statement from the authors.

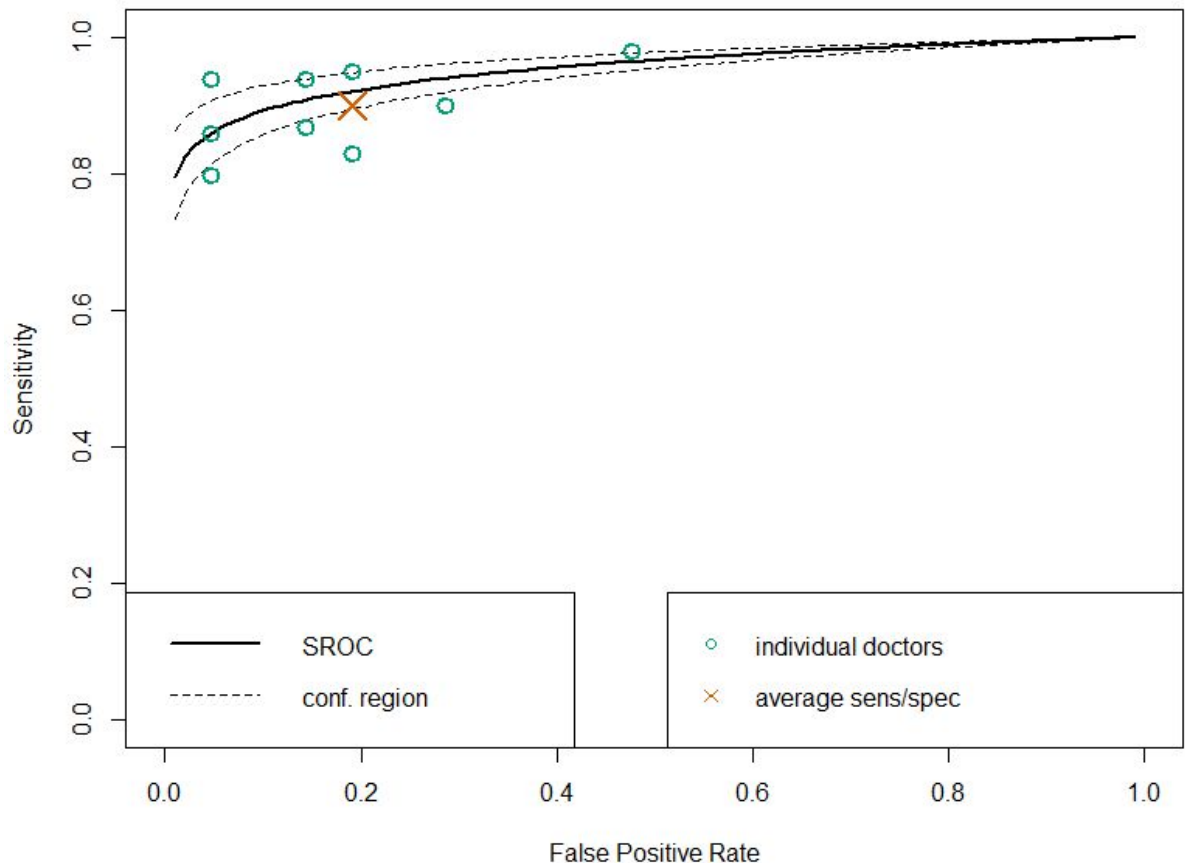

Figure 2: SROC analysis of Bien et al[22] using a random effects model, demonstrating the individual performance of doctors (green circles), the average sensitivity and specificity of doctors (orange cross) and the SROC curve (black line) with associated 95% confidence region (dotted lines).

**CheXNet: Radiologist-Level Pneumonia Detection on Chest X-Rays with Deep Learning**

Rajpurkar and Irvin et al[23] compared the performance of an AI model against 4 radiologists at the task of pneumonia detection on chest x-ray. They initially reported the average of sensitivity and specificity for the radiologists, although the primary metric was changed to the average F1 score in a later revision.

In figure 3 we demonstrate a random-effects meta-analysis of human performance. In this case we see that the average sens/spec point is quite close to the SROC curve, but the example highlights another key benefit of this approach: the confidence intervals are very wide, due to the combination of a small test dataset (with only ~60 cases of pneumonia) and the small set of readers (n = 4). By failing to account for the latter source of variation, standard statistical tests based on the average sensitivity and specificity will be biased towards the alternative hypothesis. Rajpurkar et al report that the F1 score of the model is *significantly* better than the average F1 score of the radiologist, but the meta-analytic approach suggests that this is unlikely. While we cannot perform a null hypothesis test with the information provided in Rajpurkar et al, it can be appreciated that the evidence for a meaningful difference between the model AUC (0.77, CI not provided) and the radiologist SROC AUC (0.73, 95% CI 0.66, 0.83) is not compelling.

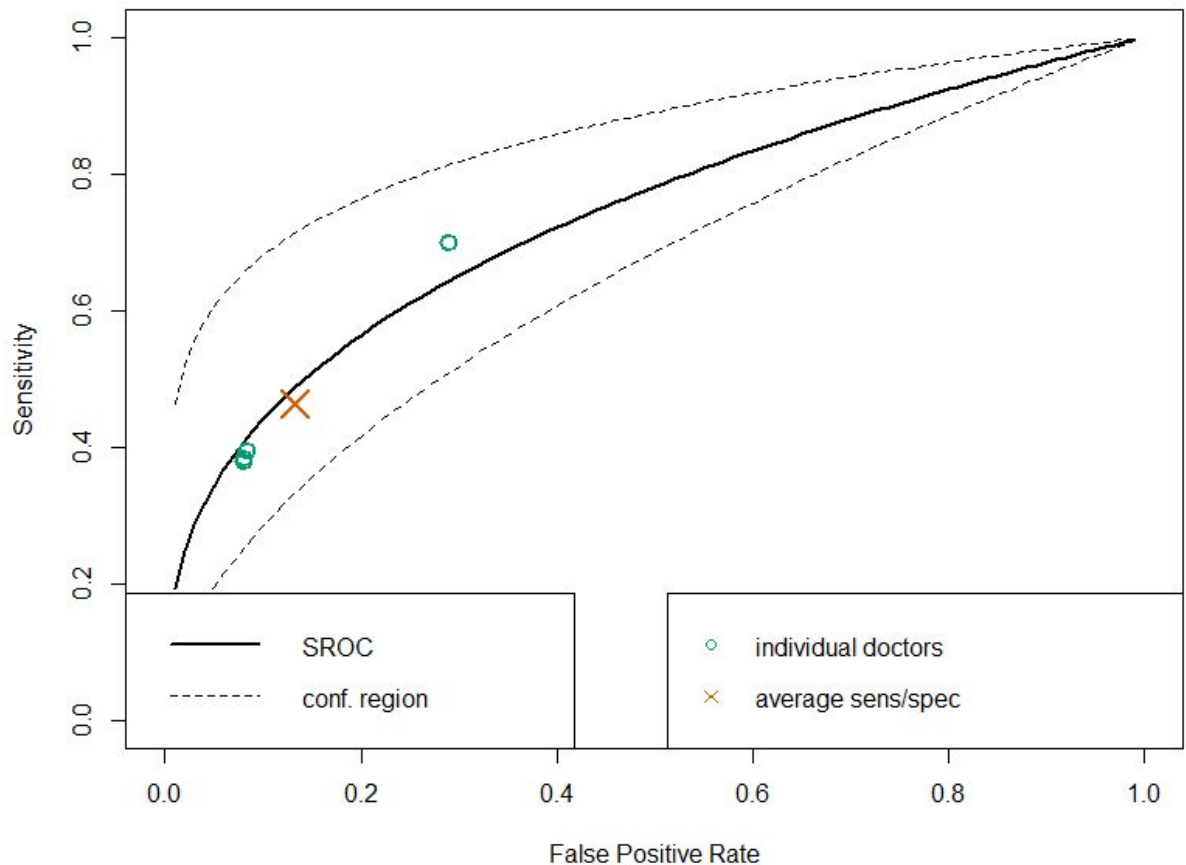

Figure 3: SROC analysis of Rajpurkar and Irvin et al[23] using a random effects model, demonstrating the individual performance of doctors (green circles), the average sensitivity and specificity of doctors (orange cross) and the SROC curve (black line) with associated 95% confidence region (dotted lines).

**International evaluation of an AI system for breast cancer screening**

McKinney et al[24] compared an AI model against radiologists for the detection of breast cancer at screening mammography. While mammography lends itself well to ordinal scoring, the authors also present results for a retrospective real-world dataset based on the binary decision of the readers with respect to the choice to perform a biopsy. Each reader read a different set of mammograms, each of different size.

In Figure 4 we demonstrate the use of a random-effects meta-analysis of human performance. The distribution of human readers is highly unusual, likely an artefact of the clinical demands of mammography (where the false positive rates of readers are monitored to standardise biopsy rates).

The SROC curve again appears to capture a reasonable "average" performance more effectively than the average of sensitivity and specificity. In this example, the average point is below the 95% CI for the SROC curve, and is biased towards the bottom right of the set of readers. Only a small number of readers, who collectively reviewed an even smaller proportion of the overall cases, are inside the average point of sensitivity and specificity.

This example demonstrates the flexibility of SROC analysis. Not only does this method appropriately manage the unusual distribution of readers, but random-effect models can estimate the variability of cases between readers, accounting for sampling bias in the setting where each reader reviews different cases.

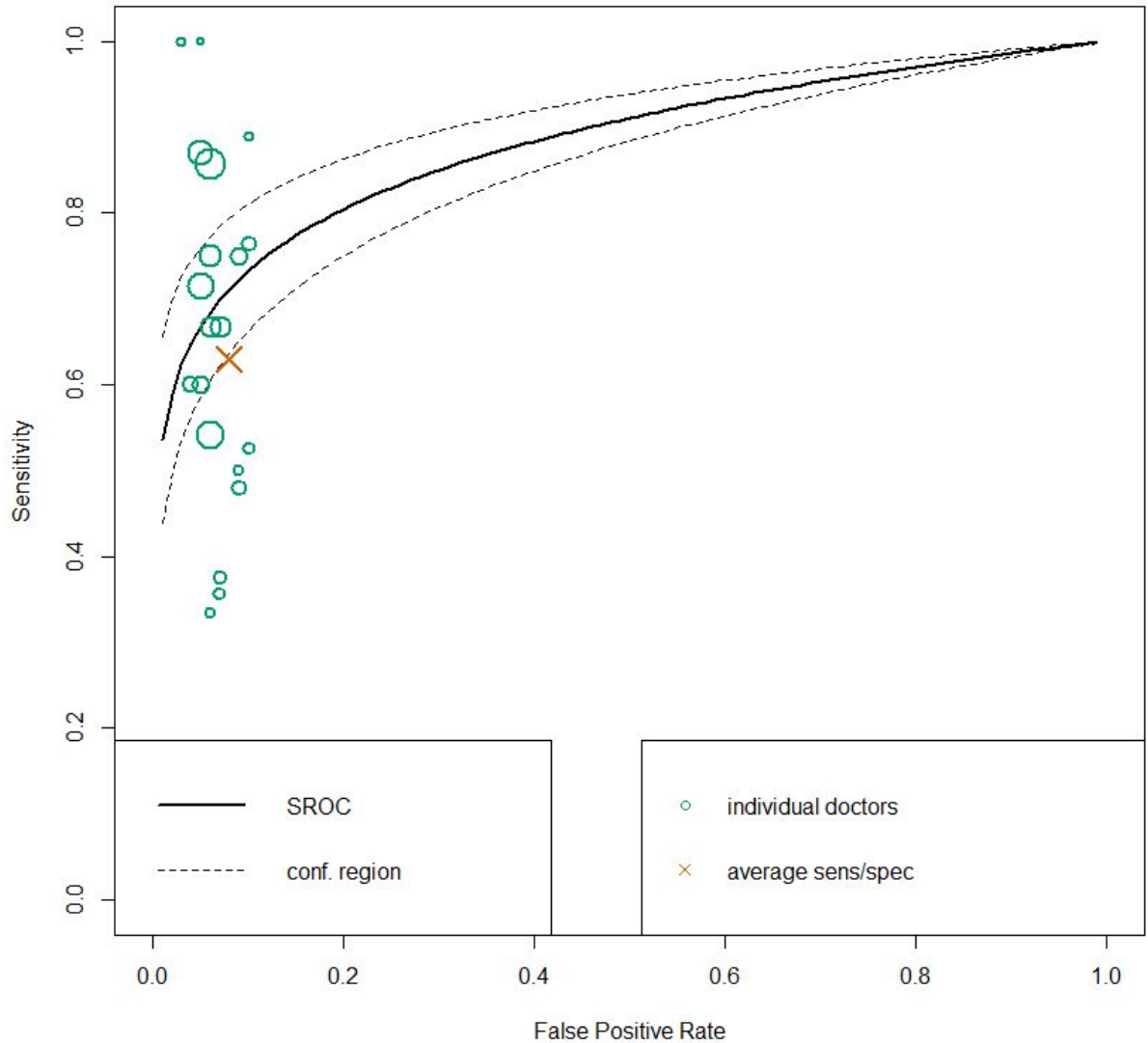

Figure 4: SROC analysis of McKinney et al[24] using a random effects model, demonstrating the individual performance of doctors (green circles), the average sensitivity and specificity of doctors (orange cross) and the SROC curve (black line) with associated 95% confidence region (dotted lines). The size of the green circles represents the number of cases each reader evaluated.

**Human–computer collaboration for skin cancer recognition**

Tschandl et al[18] report results for a 301 dermatologist reader study to classify lesions into benign and malignant categories, with each reader assessing 28 images. They report pooled average sensitivity, specificity, and several other similar statistics including the positive and negative predictive values, and the youden J statistic. Notably, there was a wide range of experience levels among the readers, ranging from less than 1 year (n = 48) up to greater than 10 years (n = 15).

The extremely large number of readers are difficult to visualise on a single plot (figure 5a), however SROC analysis can greatly improve the visibility of subgroup comparisons (figure 5b). Again, we notice that the "average" sensitivity and specificity points are well below the respective curves.

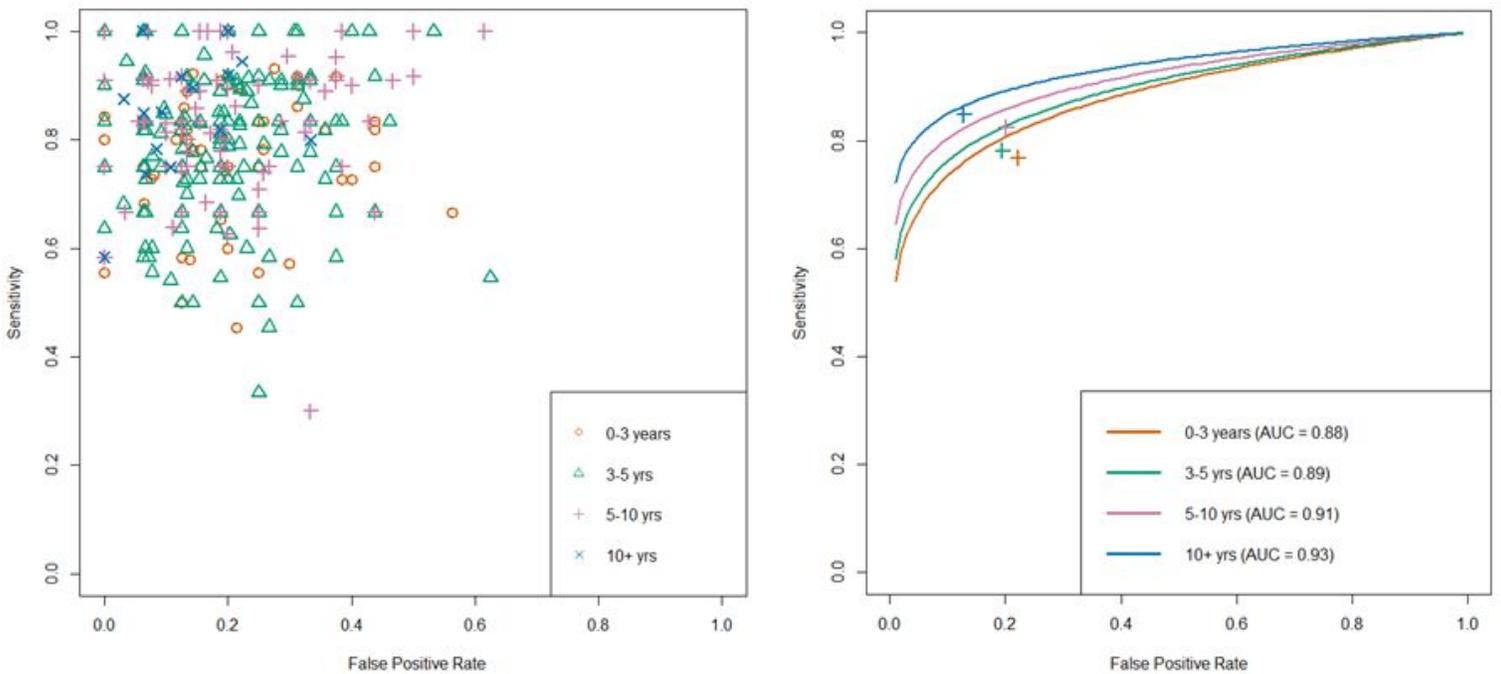

Figure 5: The individual performance of 301 human readers in Tschandl et al[18] stratified by experience level (5a, left) and summarised with SROC analysis (5b, right) using a random effects model (coloured lines) as well as the average of sensitivity and specificity (coloured crosses).

## Conclusion

The estimation of average human performance is an important application of MRMC studies, both in diagnostic specialties such as radiology and in pre-clinical studies comparing human performance with that of AI models.

In the diagnostic radiology literature, ordinal scoring systems have been widely used despite the relative lack of these in clinical practice, their biological implausibility, and the readers' lack of experience with them. In the AI literature, average human performance has been variably reported but the most common method has been to pool sensitivity and specificity independently, a technique which is methodologically flawed and will consistently bias results in favour of the AI models.

We have described the use of well validated meta-analytic techniques for the purpose of estimating average human performance where the readers produce binary diagnostic labels, and have shown the benefits of doing so by re-evaluating a number of heavily cited medical AI papers. These results show improved estimation of performance, as well as other attractive properties including providing a single metric for discrimination performance and the ability to produce estimates of variance that incorporate both the number of cases as well as the number of readers. In at least one case (CheXNet) the latter property may have altered the interpretation of a published experiment, revealing that the reported difference between human and AI performance in that work was not compelling.

These methods are not technically novel nor are they complicated, simply involving the fitting of bivariate linear models. The value of applying epidemiological meta-analytic techniques to medical AI problems arises from the availability of extensive practical experience and methodological literature regarding these techniques, the wide availability of statistical libraries to perform these operations in most common programming languages, and the flexibility of the methods. We believe that this approach can be used to standardise assessment of reader studies with binary outcomes, improving the quality and validity of these experiments in both diagnostic medicine and medical AI research.